# Search for a simultaneous signal from small transient events in the Pierre Auger Observatory and the Tupi muon telescopes

C.R.A.Augusto, V.Kopenkin[*], C.E.Navia and K.H.Tsui
Instituto de Física, Universidade Federal Fluminense
24210-336, Niterói RJ, Brazil

T.Sinzi
Rikkyo University, Toshima-ku, Tokyo 171, Japan

*Abstract*— We present results of a search for a possible signal from small scale solar transient events (such as flares and interplanetary shocks) as well as possible counterparts to Gamma-Ray Burst (GRB) observed simultaneously by the Tupi muon telescope (Niterói-Brazil, $22.9^0$S, $43.2^0$W, 3 m above sea level) and the Pierre Auger Observatory surface detectors (Malargue-Argentina, $69.3^0$S, $35.3^0$W, altitude 1400 m). Both cosmic ray experiments are located inside the South Atlantic Anomaly (SAA) region. Our analysis of several examples shows similarities in the behavior of the counting rate of low energy (above 100 MeV) particles in association with the solar activity (solar flares and interplanetary shocks). We also report an observation by the Tupi experiment of the enhancement of muons at ground level with a significance higher than $8\sigma$ in the 1-sec binning counting rate (raw data) in close time coincidence (T-184 sec) with the Swift-BAT GRB110928B (trigger=504307). The GRB 110928B coordinates are in the field of view of the vertical Tupi telescope, and the burst was close to the MAXI source J1836-194. The 5-min muon counting rate in the vertical Tupi telescope as well as publicly available data from Auger (15 minutes averages of the scaler rates) show small peaks above the background fluctuations at the time following the Swift-BAT GRB 110928B trigger. In accordance with the long duration trigger, this signal can possibly suggest a long GRB, with a precursor narrow peak at T-184 sec.

## I. INTRODUCTION

The results of search for signals from small transient events in association with the muon excess (deficit) registered at ground level (sea level) by the Tupi muon telescopes have been reported since 2005 [1,2]. Among the transient evens, there were observed signals (in association with the Fermi GBM spacecraft detector) from the interplanetary shocks as well as solar flares of small scale whose prompt X-ray emission flux is classified as C-class (flux smaller than $10^{-5}$ Wm$^{-2}$) at 1 AU [3].

The high sensitivity attained by the Tupi muon telescopes is a consequence (at least in part) of its physical location within the South Atlantic Anomaly (SAA) region ($22.88^0$ S, $43.16^0$ W). According to the Stormer's dipole approximation, the geomagnetic rigidity cutoff at the Tupi location is around 8 GV. However, in the SAA the shielding effect of the magnetosphere has a 'dip' with an anomalously weak geomagnetic field strength 22,000 nT [4] in the SAA central region ($26^0$ S, $53^0$ W), as is shown in Fig. 1. The SAA is a result of the eccentric displacement of the magnetic field center of the Earth from the geographical center (by about 400 km) as well as the displacement between the magnetic and geographic poles of the Earth. The SAA region is clearly indicated by the lowest magnetic field intensity over the Earth (Fig. 1), it is limited to the region where the magnetic field strength is less than 28,000 nT, as is shown in Fig. 1. The SAA embraces a great part of South America's central region.

Geographical distribution of proton flux measured by the HEPAD ICARE instrument on-board the Argentinean satellite SAC-C [5] shows an excess (up to 10 times) of protons with E>850 MeV in the SAA central region in comparison with the region outside of the SAA (see Fig. 2). These high energy protons are hard to be considered as Van Allen trapped protons, because the SAA models such as AP8 [5] and several measurements of the trapped protons showed that their energies do not exceed 300 MeV.

In addition, an analysis on the geomagnetic rigidity cutoff in the SAA area was made by the PAMELA collaboration [6].

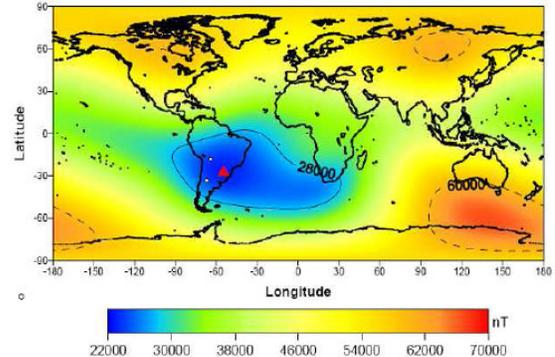

Figure 1. Geographic distribution of the geomagnetic field intensity. The SAA boundary is around B=28000 nT. The triangle indicates the SAA central region (Niterói-Brazil is near this region), where the geomagnetic field intensity is the lowest. The upper circle is the point of Chacaltaya-Bolivia and the lower point is Auger-Malargue-Argentina.

* Skobelevskaya st. 42-15, Yuzhnoe Butovo, Moscow, Russia

The magnetic rigidity of downward going particles as a function of the geographical latitude was measured with the Time of Flight (TOF) system installed at the PAMELA spacecraft. The PAMELA results show that high energy particles are present at all latitudes, while the effect of the low geomagnetic cutoff (below 1.0 GV) on low energy (E<200 MeV) particles is present only at two locations: in the region close to the poles and also in the SAA region. Downward going protons with energies above 200 MeV are also clearly seen at the latitudes between $40^0$ S and $20^0$ S (the SAA area) [6]. These are so-called 'quasi-trapped particles'. In fact, protons (ions) with energies above 300 MeV in the Earth magnetic field (the SAA area) do not satisfy the Alfven criterion for particles trapped in the magnetic field. The Alfven criterion is determined by the following relation: $r_L/r_m \ll 1$, where $r_L$ is the particle Larmor radius and $r_m$ is the curvature radius of the Earth magnetic field line. In short, the PAMELA collaboration introduced a sub-cutoff in the rigidity, that is below the nominal Stormer rigidity cutoff in the SAA area.

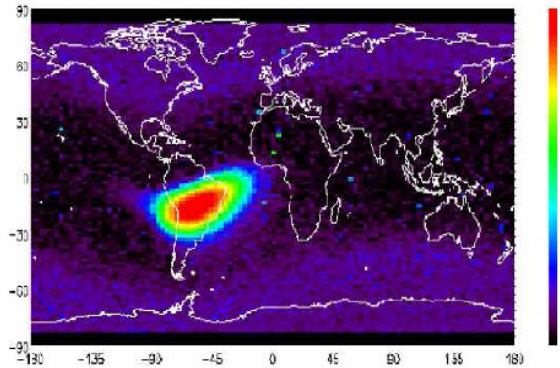

Figure 2. Geographical distribution (latitude vs longitude) of proton flux (E>850MeV) measured by the HEPAD detector [5]. The color scale is logarithmic. The red color represents 10 times higher proton flux than the one with the light blue color.

To penetrate to lower altitudes, the solar flare particles must have a rigidity above the cutoff threshold at a given latitude. As the rigidity cutoff decreases, the flux of particles (with the rigidity above the cutoff threshold at a given latitude) increases at a given altitude. The extreme cases are observed in the polar regions and in the SAA region. In these regions the solar flare induced particles will come down to very low altitudes. The solar flares observed by the Tupi telescopes also give support to this conjecture [1,2]. Thus, the shielding effect of the magnetosphere is the lowest in the SAA region. This characteristic permits the observation of small transient events such as solar flares and interplanetary magnetic shocks of diverse origins. In fact, the primary and secondary charged cosmic ray particles can penetrate deep into the atmosphere owing to the low magnetic field intensity over the SAA. Consequently, in the SAA region cosmic ray fluxes at lower energies are even higher than world averages at comparable altitudes. This is reflected as an enhancement in the counting rate of the incoming primary cosmic rays flux.

The Pierre Auger Observatory (PAO) was designed to study the physics of cosmic rays at the highest energies. So far there have been already obtained important results, such as the cosmic ray spectrum [7] and composition in the highest energy region [8]. The PAO is located in Malargue, Argentina ($69.3^0$ W, $35.3^0$ S), 1400 m above sea level and it has a Stormer's rigidity cutoff of ~9.5 GV [9]. The temporal variations in the counting rate of low-energy cosmic ray particles (muons and photons) is mainly modulated by the solar activity and properties of interplanetary magnetic field (IMF).

In the low energy region (E>100MeV) the particle flux at ground level is not constant. There are several effects, such as the local weather conditions, the diurnal solar modulation (so called ``day-night'' asymmetry). At ground the day-night asymmetry is observed as an enhancement in the particle counting rate about three hours after the sunrise (~ 12 h UT) and up to the sunset time (~ 21 h UT) at Tupi and PAO locations. The day-night asymmetry is also subject to seasonal variations. Its origin is explained by the connections between the solar and the Earth magnetic fields [10].

We show in this survey that it is possible to identify simultaneous signals from the transient solar events of small scale recorded at ground by the PAO surface detectors (SD) and the Tupi telescopes. This paper is organized as follows. In Section 2 a brief description of the PAO SD and the Tupi telescopes is given. The aim is a direct comparison between these two experiments, including such parameters as the energy thresholds and the effective solid angles. The search for the connection among spacecrafts and ground level observations (the PAO SD and the Tupi telescopes) is presented in Section 3 and Section 4. Section 3 is devoted to a search for signals associated with the solar flares of small scale, and Section 4 - for signals associated with the interplanetary magnetic shocks.

The Earth magnetic field deflects the charged particles of air showers. This deflection is caused by the component of the Earth's magnetic field perpendicular to the incident particle trajectory. As a result, there is a decrease in the number of collected charged particles that affects the detector efficiency. In the SAA region the magnetic field intensity is low, thus the capacity of the detectors to respond to incident cosmic ray particles is higher. This behavior is suitable for a search at ground of possible signals from GRB. Some results of the search are presented in Section 5, and the conclusions are drawn in Section 6.

## II. THE TUPI TELESCOPES AND THE PAO SD SCALER MODE

The PAO SD [11] is an array of more than 1600 water-Cherenkov detectors placed in a triangular grid with a spacing of 1500 m. The lateral (transverse) distribution of secondary particles of EAS at ground level is sampled by using the SD array, providing a total of about 16000 $m^2$ of collection area for the full SD array. The scaler mode consists in recording low threshold rates (scalers) using all the surface detectors of the array. The counter detectors register signals above the threshold value, corresponding to an energy of ~ 100 MeV deposited by particles that reach the detector. Since September 2005 the typical average scaler rates is around 2000 counts per second per detector. Details can be found in [12,13].

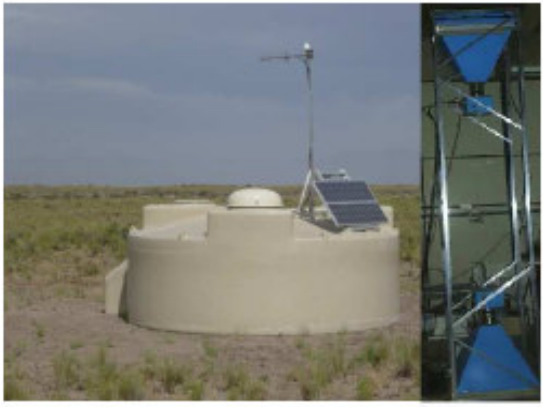

Figure 3. View of a unit of the Pierre Auger water Cherenkov detector (left) and the vertical Tupi telescope (right).

On the other hand, the Tupi experiment has two muon telescopes [14]. Each telescope was constructed on the basis of two detectors (plastic scintillators 50 cm x 50 cm x 3 cm) separated by a distance of 3 m. One telescope has a vertical orientation, and the other one is oriented near 45 degrees to the vertical (zenith), pointing to the west.

Each telescope counts the number of coincident signals in the upper and lower detector. In addition, the telescope uses a veto or anti-coincidence guard system of a third detector close to the two telescopes. This system allows only the detection of muons traveling close to the axis of the telescope. Both telescopes are situated inside a building under two flagstones of concrete, allowing registration of muons with energy threshold $E_{th} > 0.1$ GeV, required to penetrate the two flagstones. Time synchronization is essential for correlating event data in the Tupi experiment, and this is achieved by using the GPS receiver. Details of the Tupi experiment can be found in [14] and references therein.

The effective solid angle of each detector (the PAO SD and Tupi) can be roughly obtained from the following relation $\Omega = 2\pi(1-\cos\theta_z)$, where $\theta_z$ is the maximum zenith angle. For a water Cherenkov tank detector (the elementary unit of the PAO SD) the effective angle is $\theta_{eff} \sim 60^0$, that gives $\Omega_{eff} \sim 3.14$ sr. The effective field of view of the Tupi telescopes is estimated as $\Omega_{eff} \sim 0.37$ sr, around 8 times smaller than the water Cherenkov tank detector. This narrow solid angle of the Tupi telescopes is the main difference in comparison with the Cherenkov tank detector solid angle. A photograph of both detectors is shown in Fig. 3.

III. SEARCH FOR SOLAR FLARE SIGNALS AT GROUND

The Sun occasionally generates high energy (up to several GeV) particles in association with solar flares. A solar flare occurs as a result of sudden release of the magnetic energy built up in the solar atmosphere. A typical solar flare is accompanied by electromagnetic radiation. The prompt X-rays are emitted by electrons accelerated to keV energies in the explosive phase. High energy (MeV-GeV) gamma rays are produced via neutral pion decay as a result of nuclear interactions between solar protons and ions with background nuclei in the flare region [15]. The electromagnetic emission contains some information on the acceleration mechanism of particles in solar flares [16]. When the solar cycle is at its minimum, active solar regions are small and rare, so only a few solar flares are detected. The number of solar flares increases as the Sun approaches the maximum of its cycle. The period around the solar minimum (quiet conditions of the Sun) is useful for observation of small transient events, such as micro-flares.

Because of the shielding effect of the geomagnetic field, high energy (MeV-GeV) particles from solar flares can reach only high latitude regions on the Earth. The exception to this rule is possible in the SAA region, because of an area of anomalously weak geomagnetic field strength and low rigidity (low sub-cutoff) to cosmic and solar protons and ions. This feature of the SAA allows observation of small solar transient events such as solar flares of small scale.

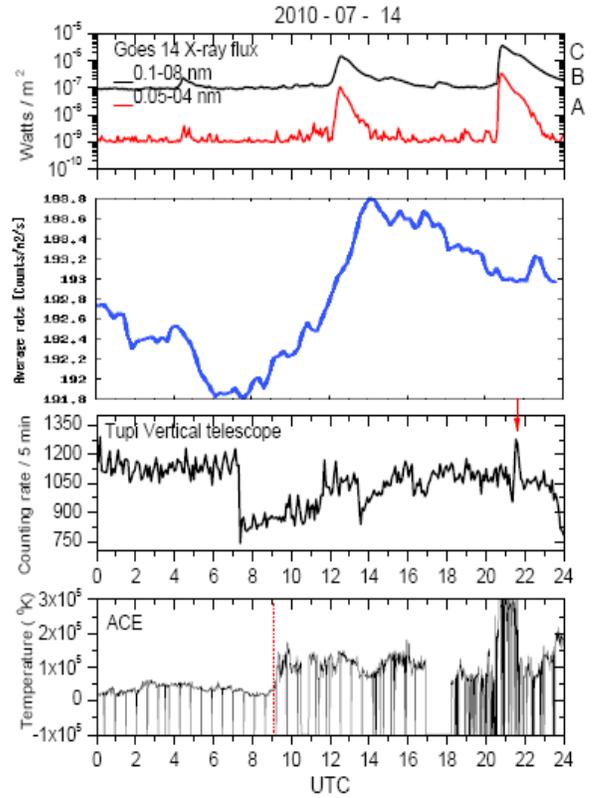

Figure 4. Example of a solar flare observation. The X-ray prompt emission of the solar flare at the onset time (20:30 UT) is classified as C-class by the GOES14 satellite (top panel). For comparisons, there are time profiles of the cosmic ray variation observed by the Earth based experiments: the PAO SD (the second panel from the top, 15 minutes averages of the scaler rates) and the vertical Tupi muon telescope (the third panel from the top, the 5-min muon counting rate). The bottom panel shows the solar wind temperature observed by the ACE spacecraft located at the Lagrange point L1. The arrow indicates the solar flare signal in the vertical Tupi telescope. Solar flares are classified as A, B, C, M or X according to the X-ray peak flux (in watts per square meter), as is shown in the top panel.

However, the detection of solar flares at ground depends on combination of several favorable conditions, related to the good magnetic Sun-Earth connection, position of the flare on the Sun disk and the onset time of the flare. For instance, if the Earth directed solar flare has been located at West of the Sun disk at the onset time approximately between (12 h UT) and (23 h UT), then it can be detected by the Tupi type detector.

On 14 July 2010 the Tupi telescopes registered a muon excess with a significance of 21% in the 5-min binning time profiles (the vertical Tupi telescope). Muon excess is associated with high energy particles emitted by solar flares (protons and ions) with energies above the pion production threshold (because they produce muons in the Earth's atmosphere),. The flare on 14 July 2010 is a nice example of an association with a solar flare of small scale registered by the GOES satellite. The X-ray prompt emission from this solar flare is cataloged by GOES as (C-class) [14]. In this survey we found that there is a possible signal associated with this flare in the PAO SD [13]. Fig. 4 illustrates the situation. The X-ray flare onset time was determined by GOES as (20:30 UT). This is approximately one hour before the muon enhancement onset time (21:27 UT) observed by the vertical Tupi telescope and about 1.58 hours before the PAO SD small peak at (22:05UT).

The bottom panel in Fig. 4 shows the time variation of the solar wind temperature as observed by the ACE-SWEPAM satellite detector, located at the Lagrange point L1. It is possible to notice that a signal in this detector is related to the same solar flare. However, the main objective of this bottom panel in Fig. 4 is to show that on 14 July 2010 there was also observed small transient event at (9:00 UT). At this time there was an interplanetary backward shock. A signal from this shock has been registered by both, the Tupi telescopes and the PAO SD. Some details of this transient event will be described in the next section.

Another example of small scale solar flare (on 27 May 2011) is shown in Fig. 5. This solar flare was observed at ground by the Tupi muon telescope. The prompt X-ray emission flux was registered by the GOES satellite at 1 AU. The flare was classified as C2.0-class. The X-ray flare onset time (14:46 UT) determined by GOES is in probable association with a muon enhancement (significance of 13%) observed by the Tupi muon telescope (15:24 UT) (see bottom and central panels in Fig. 5). This observation by the Tupi telescopes means a plausible quasi prompt emission of high energy solar protons and a coherent non-diffusive particle propagation in the interplanetary solar magnetic field. Again, in association with this small scale C class flare, there was found a signal in the PAO SD (see top panel in Fig. 5) with the onset time (15:14 UT). Thus, we can reinforce the criterion for a quasi prompt association interpretation. We see another coincidence in the position of the signals (peaks) observed by both ground based signals. This is in spite of the fact that the 5-min binning Tupi signal is sharp and narrow in comparison with the PAO SD obtained with the 15 minutes averages of the scaler rates. One can notice that there is a second flare (C2.3 class) in the data of GOES at the onset time (16:42 UT), but there is no clearly defined signal in neither the Tupi telescopes, nor the PAO SD.

In spite of low significance of the signals in the PAO SD, the identification of these two flares (14 July 2010 and 27 May 2011) in the PAO SD data was made possible through a comparison with the Tupi observation as well as the data from satellites.

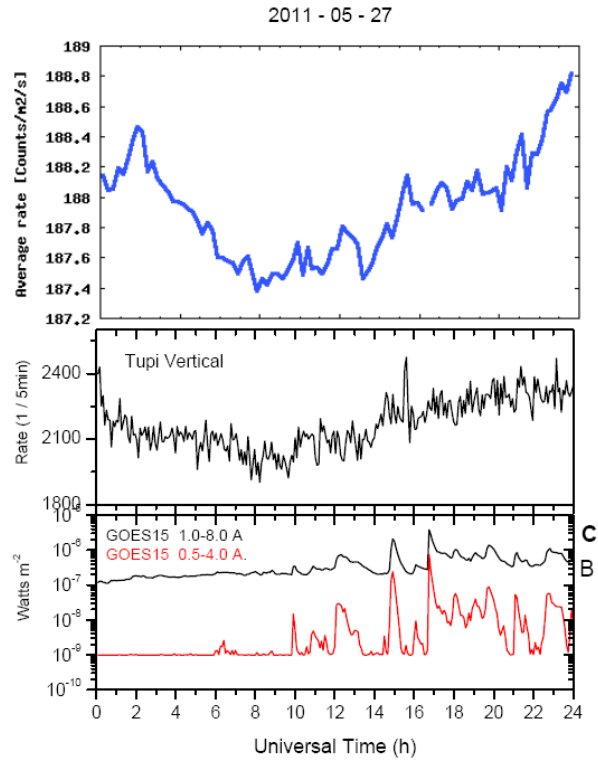

.

Figure 5. A C2.0 class solar flare signal observed at ground as particle counting rate enhancements by the PAO SD with the onset time (15:14 UT) (top panel) and the vertical Tupi muon telescope with a significance of 13% at the onset time (15:24 UT) (central panel). The GOES15 satellite X-ray prompt emission (C2.0 class) was detected at the onset time (14:46 UT) (bottom panel).

IV. SEARCH FOR INTERPLANETARY SHOCK SIGNAL

Usually interplanetary shocks are classified into two classes. The first class of the interplanetary shocks is called Coronal Mass Ejection (CME) [17]. In general, CME can be detected by the Earth instruments (as geomagnetic fluctuations) around two to three days after the CME explosion. This is the time necessary for the CME ejecta to arrive at the proximities of the Earth. If the Earth magnetic field interacts with the solar magnetic field wrapped around the 'bubble' of gas from the CME, then a fast fall in the counting rate (shadow effect) in ground detectors (for instance, neutron monitors) is observed. This effect is known as Forbush event [18]. The magnitude of the Forbush decrease depends on several factors: the size of the CME, the strength of the magnetic fields in the CME, the proximity of the CME to the Earth, and the physical location of the detector. Search for muon enhancement at sea level from CME was reported elsewhere [3].

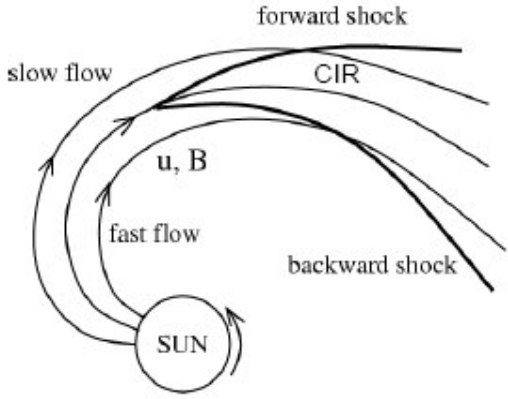

Figure 6. An illustration of CIR formation mechanism. The high speed solar wind, originating in coronal holes, interacts with the preceding slow solar wind forming forward and reverse shocks. Marks are: u - speed of the solar wind, B - magnetic field strength.

The second class of the interplanetary shocks consists of corotating interaction regions (CIRs) [19]. These are regions of compressed plasma formed at the leading edges of corotating high-speed solar wind streams originated in coronal holes, as they interact with the preceding slow solar wind. The CIRs are particularly prominent features of the solar wind during the declining and minimum phases of the 11-year solar cycle. Fig. 6 shows the mechanism of CIR formation, where the forward shock and reverse shock are the regions where particles of interplanetary space are accelerated .

The registration of shock waves at ground in association with the spacecraft detectors located at the Lagrange point L1 allows to determine whether the shock was propagating forward or backward (in relation to the Sun). The forward shock configuration happens if the CIR is formed at a distance less than 1 AU, particularly when the CIR spread is close to the Sun-Earth direction. In this case, at first the shock reaches the Lagrange point L1 and some time later it arrives to the Earth.

On the other hand, registration at ground of backward shock waves (directed away from the sun) is possible if the CIR is formed at a distance above 1 AU. In this configuration, at first the shock reaches the Earth and then the Lagrange point L1. Depending on the shock speed, there can be expected a time delay of the shock (for about 1-2 hours) between the observation of the shock at the Lagrange point L1 and the Earth, or vice versa.

Example of a CIR observed at the distance of 1 AU on 14 July 2007 at (7:15 UT) as a fall in the Tupi muon counting rate is shown in Fig. 4 (see previous section). In this interpretation, the shock wave crossed the Earth at first, and then, about 1.6 h later (see vertical lines) it reached the spacecraft ACE position, that is located at the Lagrange point L1. There is observed increase in the temperature of the solar wind (see Fig. 4). Consequently, this shock can be classified as a reverse shock. A clear association between the Tupi signal (a fall in the counting rate with 37% significance) and the ACE satellite solar wind temperature is observed. A signal also is found in the PAO SD. This signal looks as a fall in the counting rate (~18% significance ) with the onset time ~ 1 hour in relation to Tupi.

On 29 March 2011, according to the ACE and SOHO solar detectors, there was an interplanetary forward shock with the onset time (15:00 UT) with a confidence level of 68%. The effect of this shock was detected by ground detector (the Tupi muon telescope) as a sudden increase followed by a fall in the counting rate (shadow effect). There is ~ 1 hour time delay between the onset time determined by the satellites and the Tupi telescopes. The counting rate variation in the Tupi telescope has ~ 10% significance. A signal is observed in the PAO SD. There is a fall in the counting rate with a significance near 1 %. Fig. 7 illustrates the situation.

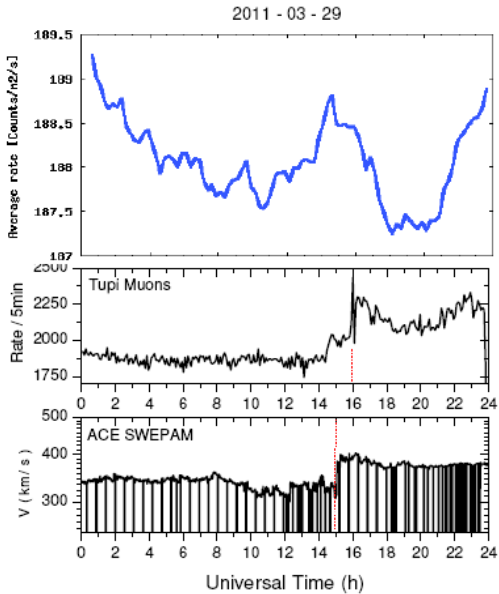

Figure 7. Example of a forward interplanetary shock (CIR) observed at 1 AU as a fast rising in the muon counting rate registered by the inclined Tupi telescope on 29 March 2011 (central panel). The signal observed by the PAO SD is shown in top panel. The solar wind speed observed by the ACE spacecraft at Lagrange point L1 is presented in bottom panel. The vertical lines indicate the CIR onset time determined by Tupi and ACE. A sudden increase followed by a fall in the counting rate (shadow effect) can be seen in the ground based data.

V. SEARCH FOR GAMMA RAY BURST SIGNAL

Gamma Ray Bursts (GRBs) are probably the most powerful explosions observed in the Universe today. The total energy released is typically of the order $10^{51}$ erg. The origin and production mechanism of GRB is not yet understood very well. However, there are some evidences that they are probably generated by collapsing massive stars (for long GRBs) and merging of two neutron stars (for short GRBs). Some long GRBs are usually followed by emission of all sort of radiation from radio waves to GeV photons. The GRB sources are randomly distributed over the whole sky, and the GRBs are observed by satellite based detectors at a rate as much as 2-3 bursts per day. Information from GRBs by the satellite gamma ray detectors are distributed to ground based detectors via the GRB Coordinate Network (GCN) [20] . The main parameters are the GRB trigger time and the GRB trigger coordinates. They are essential for a search of counterparts to GRB at ground (GeV component as well as afterglow). In March 2005 the scaler mode was implemented in all the detectors of the PAO SD [21], in order to search for GRBs. It was also included in the LAGO project [22], a network of water Cherenkov detectors at high mountain altitudes. So far, no positive results have been reported by either of these detectors.

However, the Tupi observations have produced some experimental evidences for a delayed signal in relation to the Swift GRB 080723 and Fermi GBM 081017474 [23]. In addition, the event Konus GRB090315 (Hurley et al., GCN Circular 2009) and the Swift GRB091112 [24] have probable correlations with the observed muon excess.

On 28 September 2011, a sharp peak with a significance above $8\sigma$ was found in the 24 hours raw data (counting rate at every second) of the vertical Tupi telescope. It was possible to recognize this peak in the time profile of the muon excess just by naked eye (see Fig. 8). The signal was absent in the inclined Tupi telescope. The peak with a duration of one second has been registered in close temporal coincidence (T-184 sec) with the Swift-BAT GRB110928B (trigger=504307).

that the Swift-BAT GRB110928B at the position of (RA, Dec) = (278.950, -19.328; J2000) [25] corresponds to the MAXI source J1836-194, lying only 70 arc-seconds away [25,26], where 'MAXI' stands for a sensitive X-ray slit camera attached on the Japanese Experiment Module (JEM) Kibo at the International Space Station. It is interesting that the MAXI J1836-194 is the newly discovered (on 2011 August 30) transient source at the position of (RA, Dec) = (279.12,-19.41; J2000) in the Sagittarius constellation [27].

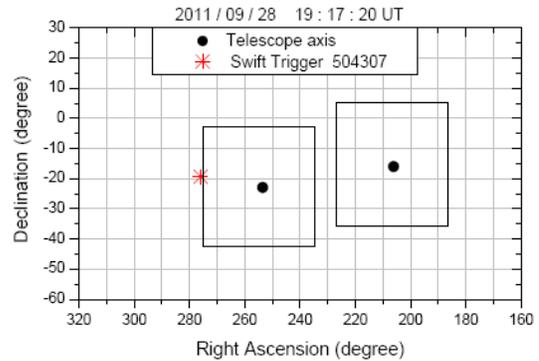

Figure 9. The equatorial coordinates of the Tupi telescopes (vertical, on the left and inclined, on the right) axes (black circles). Squares represent the field of view of the telescopes and the asterisk is the position (coordinates) of the Swift-BAT GRB110928 (trigger=504307).

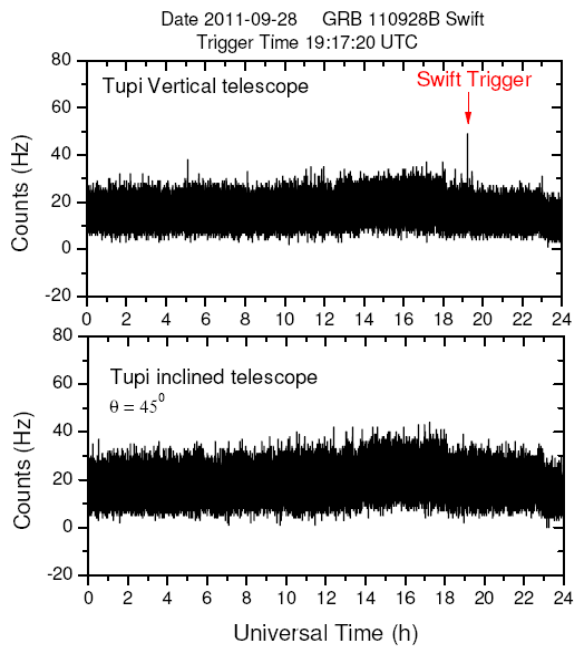

Figure 8. Time profile of the 1-sec binning counting rate (raw data) observed by the vertical (top panel) and inclined (bottom panel) Tupi telescopes on 28 September 2011. The vertical arrow indicates the Swift-BAT GRB 110928B trigger.

The Swift-BAT trigger at (19:17:20 UT) had a long duration of 17.1 min [25], and the image significance was determined as $7.79\sigma$ [25] . Fig. 8 summarizes the situation, where the time profiles of the vertical and inclined Tupi telescopes on 28 September 2011 are shown. It was reported

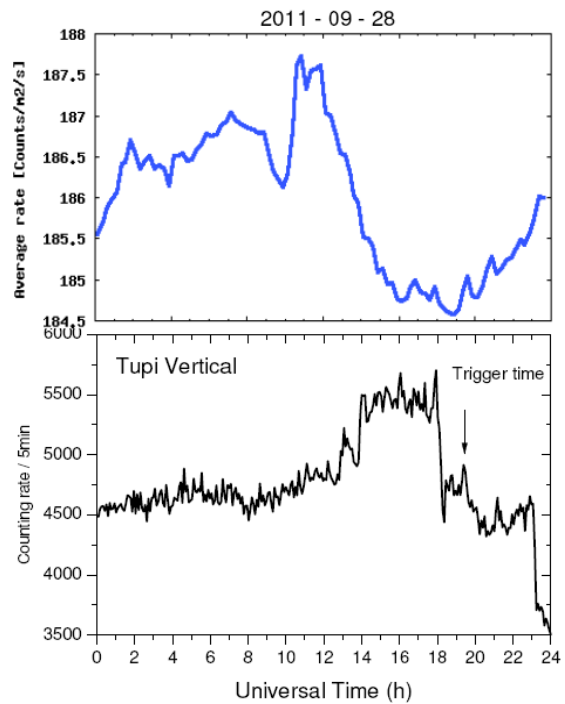

Figure 10. Comparison between the time profiles observed on 28 September 2011 by the PAO SD (top panel) in the 15 minutes averages data and the 5-min muon counting rate in the vertical Tupi telescope (bottom panel) on 28 September 2011. The vertical arrow indicates the time of the Swift-BAT GRB 110928B trigger.

It was also reported as a black hole candidate X-ray binary [28]. It is necessary to stress that the trigger coordinates are in the limit of the effective field of view of the vertical Tupi telescope (as can be seen in Fig. 9).

We searched for a possible simultaneous signal in the PAO SD as well. The 5-min muon counting rate in the vertical Tupi telescope as well as publicly available data from the PAO SD (15 minutes averages of the scaler rates) show small simultaneous peaks above the background fluctuations at the time following the Swift-BAT GRB 110928B trigger. If confirmed, this signal could possibly reinforce the GRB interpretation, as is shown in Fig. 10.

Evidently, a more robust analysis in the original PAO SD is necessary, in order to confirm whether this small signal is really a true signal, or just a mere coincidence with some background fluctuation. As we do not have access to the PAO SD raw data, the fine analysis was made only on Tupi data.

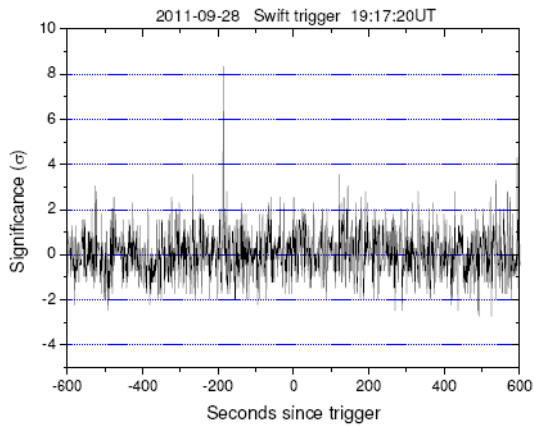

Figure 11. Statistical significance (number of standard deviations) of the 1-sec binning counting rate observed by the vertical Tupi telescope as a function of the time elapsed since the Swift-BAT GRB110928 trigger time.

In order to see the background fluctuation structure, all the time bins (each of 1 sec duration) of the vertical Tupi telescope have been tested by Bin Selection Criteria (BSC) [10]. According to this algorithm, the statistical significance in the i-th bin is defined as

$$\sigma_i = \frac{(C^{(i)} - B)}{\sqrt{B}},$$

where $C^i$ is the measured number of counts in the bin and B is the average background count. This function follows a Gaussian distribution if there is no signal. The results are shown in Fig. 11. The signal was detected 184 sec before the Swift-BAT GRB 110928B trigger. It is possible to see that the signal in the vertical Tupi telescope has a significance above $8\sigma$.

To see the structure of the counting rate, a confidence analysis of the background was made. Initially, we have used the complete 24 h raw data of the vertical telescope for 28 September 2011. We have also chosen the interval of one hour around the Swift-BAT GRB110928 trigger time. The significance distribution of the muon counting rate in the vertical telescope is consistent with a Gaussian distribution, as is shown in Fig. 12. The trials with a confidence of $8.3\sigma$ correspond to the peak at (T-184 sec) from the Swift-BAT GRB110928B trigger (trigger=504307).

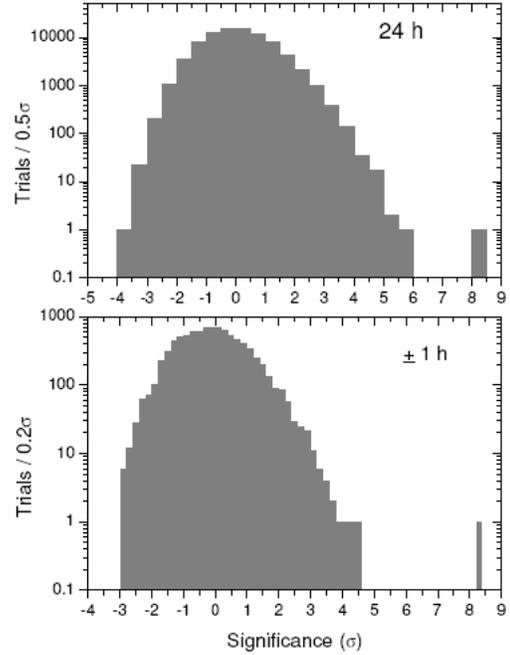

Figure 12. Distribution of the fluctuation counting rate for the vertical Tupi telescope (in units of standard deviations). Top panel: for a period of 24 hours (28 September 2011). Bottom panel: using a time windows inside of one hour interval around the Swift-BAT GRB110928 trigger.

In addition to that, there were found no flare or transient event (see Fig. 13), as well as no anomalous changes in the atmospheric pressure, temperature or other known environmental conditions during the time period close to the signal detection on 28 September 2011. One can notice that the width of the peak at the half of the height (near the Swift trigger) is close to the trigger duration (see Fig.13).

## VI. CONCLUSION

It is very well-known that the primary charged particles come deep down into the atmosphere owing to the low field intensity over the SAA. Consequently, cosmic ray fluxes at low energies in the SAA region are even higher than world averages at comparable altitudes. This situation is reflected as an excess in the counting rate of the primary cosmic rays. Probably this behavior is responsible for the observation of muon excess in association with high energy particles (protons and ions) emitted by flares of small scale, as has been reported by Tupi experiment.

The flux of particles at ground (E>100 MeV) is not constant, because they are subject to temporal variations due to local weather conditions as well as they are modulated by the solar activity and fluctuations of interplanetary magnetic field. We have presented in this survey a search for a simultaneous signal of solar transient events of small scale in the PAO SD and the Tupi muon telescopes. Both experiments are situated within the SAA area. Of several candidates that were found we presented two events associated with solar flares and two events associated with interplanetary shocks (CIR).

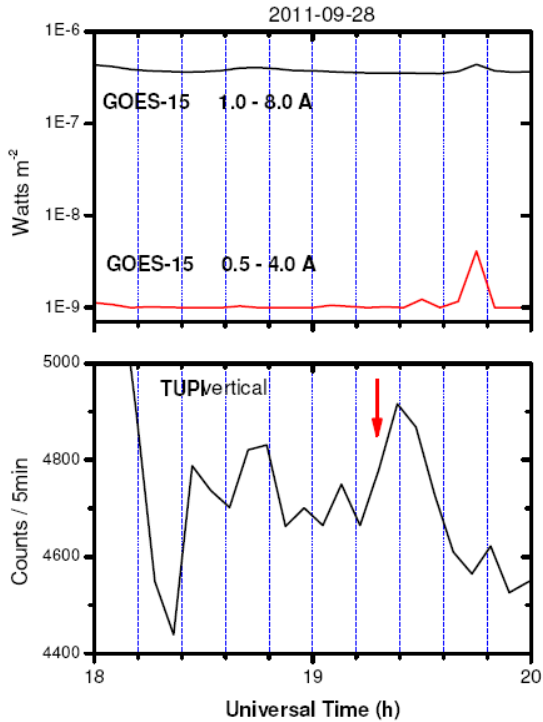

Figure 13. Top panel: The x-ray flux on 28 September 2011, according to GOES 15, for two wavelengths. Bottom panel: The 5-min muon counting rate in the vertical Tupi telescope. The vertical arrow indicates the time of the Swift-BAT GRB 110928B (trigger=504307).

We searched also for possible coincident counterparts to the GRB signals in the narrow field of view of the Tupi muon telescopes and in the large solid angle water Cherenkov detectors of the PAO SD. We found a candidate event with features of likely association with the gamma ray burst GRB 110928B observed by Swift satellite. The light curve, including background duration and rate significance are not determined. However, the GRB has an image significance $7.79\sigma$ and the trigger with duration of 17.1 min [25]. In addition, the burst location was close to the MAXI source J1836-194 [26].

The GRB observation in the limit of the effective field of view of the vertical Tupi telescope consists of an excess of muons with a significance higher than $8\sigma$ in the 1-sec binning counting rate of muons (raw data) in close time coincidence (T-184 sec) with the Swift-BAT GRB 110928B trigger (trigger=504307). The 5-min muon counting rate in the vertical Tupi telescope as well as publicly available data from Auger (15 minutes averages of the scaler rates) show small simultaneous peaks above the background fluctuations at the time following the Swift-BAT GRB 110928B trigger. It suggests a long GRB, with a narrow peak 184 sec before the Swift trigger time. The observations at multi-wavelengths by various detectors are specially interesting for the previously unknown sources. Evidently, in order to confirm whether this small signal of low significance in the PAO SD is really a true signal, or a mere coincidence with a background fluctuation (several other peaks of this type can be found in the Auger data), a more robust analysis of the original PAO SD is necessary. While the signal in the vertical Tupi telescope seems to be a good candidate for the counterpart associated to the GRB 110928B, a precise interpretation of the Tupi data on the basis of Monte Carlo simulations is necessary, in order to obtain the fluence associated with this mysterious GRB candidate. This will be the next step.

ACKNOWLEDGMENT

This work is supported by the National Council for Research (CNPq) of Brazil, under Grant 306605/2009-0 and 01300.077189/2008-6 and Fundação de Amparo a Pesquisa do Estado do Rio de Janeiro (FAPERJ), under Grant 08458.009577/2011-81 and E-26/101.649/2011. We are also grateful to various catalogs available on the web and to their open data policy, especially to Pierre Auger Observatory (event display), the ACE Science Center and the GCN Catalog.